\documentclass[a4paper,12pt]{article}
\usepackage[cp1251]{inputenc}  %Cyrillic Windows
\usepackage[english]{babel}
\usepackage {euscript}
\usepackage{epsfig}
\usepackage{amsmath}
\usepackage{amssymb}
\usepackage{bm}
\usepackage{indentfirst}

\newcommand{\beq}{\begin{equation}}
\newcommand{\eeq}{\end{equation}}
\newcommand{\beqn}{\begin{eqnarray}}
\newcommand{\eeqn}{\end{eqnarray}}

\begin{document}

\begin{center}
{\bf \large Positron production in collision of heavy nuclei}
\end{center}

\begin{center}
I.B. Khriplovich\footnote{khios231@mail.ru}
\end{center}
\begin{center}
Department of Physics, St. Petersburg State University, 7/9 Universitetskaya Nab., 
\\ St. Petersburg 199034, Russia
\end{center}

\bigskip

\begin{abstract}

We consider the electromagnetic production of positron in collision of slow heavy nuclei, 
with the simultaneously produced electron captured by one of the nuclei. The cross-section 
of the discussed process exceeds essentially the cross-section of $e^+ e^-$ production.

\end{abstract}


\vspace{8mm}

The positron production in collision of slow heavy nuclei (as well as the production of $e^+ e^-$ pair), was addressed in numerous
papers (see, for instance, [1 - 3]). In the present note we reconsider the problem of the 
positron production in collision of heavy nuclei, with the electron captured by another 
nucleus. We present a simple direct solution of this problem.

\begin{figure}[h!]
\centering
\includegraphics[width = 0.5\textwidth]{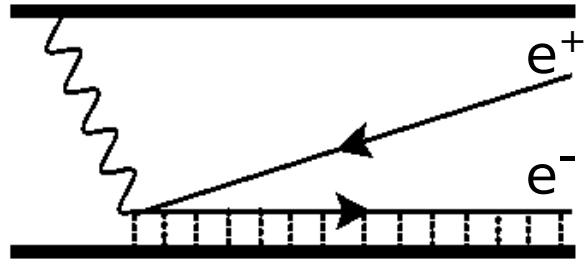} 
\caption{
Feynman diagram of positron production.
}
\label{fig:1a}
\end{figure}

The process under discussion is presented in Fig.~\ref{fig:1a} 

The heavy lines in the Figure refer to the propagation of nuclei. The wavy line refers 
to the virtual photon. In the produced pair $e^+ e^-$, the electron is captured immediately 
by one of the nuclei. As to the positron, we assume that its interaction with the nuclei 
can be neglected. Simple symmetry arguments demonstrate that this repulsive interaction 
cannot be strong.

Then, it can be easily demonstrated that diagram Fig.~\ref{fig:1a} is the only one which 
survives in the limit of heavy mass $M \to \infty$ of colliding nuclei.

Let us come back now to Figure 1. 
The electromagnetic vertex on the upper heavy line reduces to $(p^\prime + p)/2M \approx v$ 
(we neglect here the difference between $p^\prime$ and $p$). With the amplitude proportional 
to $v$, the discussed total cross-section is proportional to v also:
\beq
v^2/v = v;
\eeq
$v$ in the denominator of (1) originates from the particle flux. For the velocity of nuclei, 
we assume $v \simeq 0.1$.

Thus, we can confine to diagram 1, with the overall factor $Z^2 \alpha^2$ in the cross-section. 
One factor  2 in this expression arises since the electron can be captured by any of the two nuclei. 
One more factor 2 is due to two polarizations of the produced positron.

The total cross-section is obviously a scalar, and its essential part is $1/m^2$.  

In this way, we arrive at the following result for the total cross-section of the positron production:
\beq
\sigma = 4\,Z^2 \alpha^2 \frac{v}{m^2}.  
\eeq

Numerically, this cross-section is

\beq
\sigma \simeq 10^{-22} \,\rm{cm^2}.
\eeq

This cross-section exceeds essentially the cross-section of $e^+ e^-$ production in collision of the uranium nuclei, which is close to $10^{-25}$ cm$^2$.

\subsection*{Acknowledgements}

I am grateful to V.M. Shabaev  for useful discussions.

\renewcommand{\bibname}
\end{document}